# Nearsightedness of Electronic Matter and the Size of Viruses

W. T. Geng[a]

*School of Materials Science & Engineering, University of Science & Technology Beijing,*

*Beijing 100083, China,*

May 11, 2010

I conjecture that the nearsightedness of component electronic matter largely determines the size of a virus. These two length scales, one from physics and one from biochemistry, are in fact the same dimension which connects our quantum and everyday worlds. Learning how viruses interact with microscopic molecules and macroscopic biological cells might help us understand the quantum-to-classical transition in general cases of multiscale phenomena.

[a] E-mail: geng@ustb.edu.cn

The universe, perceived by we human beings, presents itself in multiples scales. Electrons, atoms, and molecules are perfectly described by quantum mechanics, whereas classical physics works well in our everyday life. We are sure that at a certain length scale quantum-to-classical transition (QCT) is bound to happen; yet its implication beyond physics is not very clear. In the advent of high performance supercomputers, it is now feasible to deal with hundreds or even thousands of atoms fully quantum mechanically. Prototypical multi-scale phenomenon such as cracking in solid materials is about to be within reach of quantum mechanical computations. In fact, details of the atomic and electronic structure near the tip of a crack in both insulating[1] and metallic[2] materials are being revealed, and the calculations are expected to expand to the classical (elasticity) region soon. Another example of quantum-classical boundary in physical virology is the nanoindentation of phage capsid.[3,4] Perhaps within a decade we will be able to simulate the penetration of RNA or DNA of a virus through a protein membrane using fully quantum mechanical computations. What can we learn about the QCT from these first-principles calculations when someday they become achievable?

We can argue that such a QCT is closely related with the nearsightedness of electronic matter (NEM)[5]. NEM is a many-body effect in the quantum physics realm and was realized and introduced by Kohn[6] in the year 1996 and was sharpened and quantified by Prodan and Kohn in the year 2005 [Ref. 5]. For electronic matter, this nearsightedness principle declares that for a given unperturbed system and a given $R$, the changes of electron density at $r_0$ due to any perturbing potential, $\Delta n(r_0)$, have a finite maximum magnitude, $\overline{\Delta n}$, which decays monotonically as a function of $R$, $\lim_{R \to \infty} \overline{\Delta n}(r_0, R) = 0$. With NEM, materials of atomic nature manifests itself in an additive or averaged manner and any atomic perturbation or boundary effect therein shall fade away along with distance when it goes from micro- to macro-scale. As a consequence, a piece of electronic matter of finite size can be viewed as arbitrarily large in view of its chemical properties, and hence a classical world in our naked eyes. QCT is nothing new, but it poses a fundamental question: Is there a characteristic QCT length scale? Or, in other words: is there a *clear* border separating the quantum and classical worlds? By *clear* we mean this characteristic length scale depends at most only on the basic electronic properties of this material.

Here I recall a well-known phenomenon in metallurgy. Near the grain boundary in a polycrystalline alloy, often there is an aggregation or depletion of some impurities or alloying additions due to perturbation of the periodic crystal field by the grain boundary.[7] Figure 1a [Credit: Ref. 7] displays the depth dependence of grain boundary segregation of various impurities in a number of alloys. It is clear that at lattice sites over 20 Å away from the boundary, the impurity content recovers well to the bulk value. A more recent and more accurate measurement using three-dimensional atomic tomography technique further confirms the above observation.[8] In Figure 1b [Credit: Ref. 8] we see that except for carbon, which has a very low concentration, molybdenum, phosphorus, and boron all show significantly a segregation depth of about 20 Å near the grain boundary in Ni-Fe alloy. These measurements demonstrated unambiguously that the nearsightedness of many a kind of metallic systems is about 20 Å.

**Fig. 1 Depth dependence of grain-boundary segregation. (a) [Credit: Ref. 7, segregation of S, P, Sn and Sb at the grain boundary in various alloys. (b) [Credit: Ref. 8, segregation of C, P, B, and Mo ] at the grain boundary in Ni-Fe alloy.**

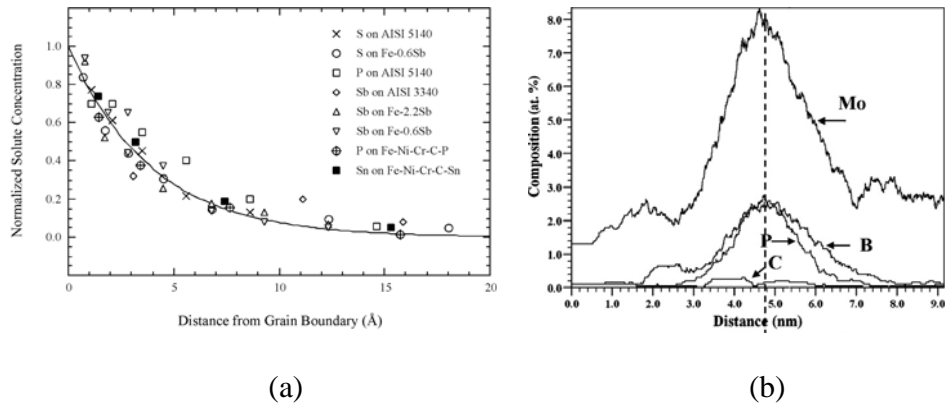

(a)                                          (b)

In his famous lectures, Schrödinger[9] put it clearly that "*The working of an organism requires exact physical laws.*" Taking this statement as a fundamental law of biology, it is reasonable to assume that in order to minimize the quantum effect on the biological body as a whole, a living thing made of electronic matter has to be considerably larger than the sightedness, or QCT length, of the condensed matter thereof. Well below the QCT length scale, there are atoms and molecules which are all non-living things obeying quantum mechanics. Well above the QCT length, there are clay, silt, sand and larger objects whose equation of motion are well depicted by classical mechanics. At the QCT length scale, in between the micro- and macro-worlds and also between living and non-living things, we have viruses. Viruses are electronic matter entities that are really mesoscopic. They exhibit parts of characteristics of both non-living and living things. Viruses of the same kind can form crystal at low temperature, just as identical molecules. On the other hand, their capability of reproduction and evolution are also typical features of living things.

I propose that *viruses must be large enough to allow deterministic physical laws, yet small enough to keep indistinguishability for eternal life*. Only with indistinguishability can a particle keep its exact structure for an infinitely long time; and only with a non-microscopic size, can a virus fight against quantum uncertainty and reproduce itself without making too many mistakes. Mimivirus, discovered in 1992, has a capsid diameter of 4000 Å (on the same order of the wavelength of visible light), is the largest virus found ever. One of the thinnest viruses is *bacteriophage fd*, which is rod-like and has a diameter of 6.6 Å, being just larger than the sightedness of metallic materials and capable of minimizing the boundary effect. If the component material is insulating, its sightedness can be much larger depending on the energy gap separating the occupied (HOMO) and unoccupied (LUMO) electronic state. Therefore, I anticipate the thinnest rod-like viruses to be metallic in the short dimension, and insulating along the rod. If a virus is too small to surpass the sightedness of its component electronic matter, it cannot carry well-defined physical quantity in classical sense such as volume, bulk and bending modulus to make exact physical laws applicable; if it is too large, however, it will lose its indistinguishability and cannot keep its characteristic atomic structure for an extremely long time.

Viruses are not only the bridge connecting non-living and living things, but more importantly, in the eye of physicists, are also the bridge connecting micro- and macro-scale systems. Therefore, we might take the size of viruses as the characteristic length of QCT. In view of the nearsightedness of electronic matter induced by quantum many-body effect, we can understand why viruses have their present size. Modifying the HOMO-LUMO gap of a virus by chemical or physical adsorption could be an effective way to disable it. Moreover, studies on viruses could be the biology we need for physics,[10] shedding light on our understanding of the quantum-to-classical transition in general and hence many intriguing phenomena manifesting themselves in multiple scales.

---

[1] J. R. Kermode1, T. Albaret, D. Sherman, N. Bernstein, P. Gumbsch, M. C. Payne, G. Csányi & A. De Vita, Nature **455**, 1224 (2008).

[2] W. T. Geng, arXiv:1001.3168 (2010).

[3] M. Buenemann and P. Lenz, PNAS, **104**, 9925 (2007).

[4] W. H. Roos and G. J. L. Wuite, Advanced Materials, **21**, 1187 (2009).

[5] E. Prodan, and W. Kohn, PNAS, **102**, 11635 (2005).

[6] W. Kohn, Phys. Rev. Lett. **76**, 3168 (1996).

[7] M.P. Seah in *Practical Surface Analysis by Auger and X-ray Photoelectron Spectroscopy*, edited by D. Briggs and M.P. Seah, (1983) 247-282.

[8] D.H. Ping, Y.F. Gu, C.Y. Cui, and H. Harada, Materials Science and Engineering A **456** (2007) 99–102

[9] E. Schrödinger, *What is life?* Cambridge University Press, Cambridge, U.K. 1944.

[10] Here we do not mean *biophysics* which stands for the physics for biology, nor *physical biology* which stands for the biology described using physics.